\documentclass{optica-article}

\journal{oe}
\usepackage{makecell}
\usepackage{lineno}
\usepackage{pifont}

\begin{document}


\title{Multiple plateaus of high-sideband generation from Floquet matters}

\author{ Yu-Xuan Chen,\authormark{1} Gan Wang,\authormark{1} Mingjie Li,\authormark{3,4,5} Tao-Yuan Du\authormark{1,2,*}  }

\address{\authormark{1}School of Mathematics and Physics, China University of Geosciences, Wuhan 430074, China\\
\authormark{2}Key Laboratory of Computational Physical Sciences (MOE), Fudan University, Shanghai 200433, China\\
\authormark{3}Department of Applied Physics, The Hong Kong Polytechnic University, Hung Hom, Kowloon, Hong Kong, China 

\authormark{4} Shenzhen Research Institute, The Hong Kong Polytechnic University, Shenzhen, Guangdong, 518057, China

\authormark{5} ming-jie.li@polyu.edu.hk }

\email{\authormark{*}Corresponding author: duty710@163.com} 

\begin{abstract}	
We theoretically report that high-order sideband generation (HSG) from Floquet matters driven by a strong terahertz light while engineered by weak infrared light can achieve multiple plateau HSG. The Floquet-engineering systems exhibit distinctive spectroscopic characteristics that go beyond the HSG processes in field-free band-structure systems. The spatial-temporal dynamics analyses under Floquet-Bloch and time-reversal-symmetry theories clarify the spectral and its odd-even characteristics in the HSG spectrum. Our work demonstrates the HSG of Floquet matters via Floquet engineering and indicates a promising way to extract Floquet material parameters in future experiments.
\end{abstract}.

\section{Introduction}

The interaction between light field and matter can regulate the dynamic process of Floquet materials \cite{Oka,Linder,Torre}, with particular emphasis on modifying and controlling quantum states through periodic light fields, known as Floquet engineering \cite{Rudner1,Minguzzi,Rudner2,Machado,Rechtsman}. This method shows promise in tailoring the physical properties of Floquet materials due to the existence of light-dressed band structures, facilitating exploration into non-trivial phases in topological insulators. Many recent experiments have also demonstrated the practical feasibility of such a Floquet engineering \cite{McIver,Kobayashi,Zhou,Wang,Ito}. The promising nature of Floquet engineering inspires us to explore the underlying superiority of the aspect of the generation of strong-field coherent light sources.

Under the driving of the periodic light field, the process of high-harmonic generation (HHG) constitutes a particular Floquet process, which is observable across both perturbative and non-perturbative generation regimes. HHG could provide a unique table-top source of coherent emission in extreme UV and X-ray spectral regions and serves for producing attosecond pulses for ultrafast spectroscopy \cite{Lewenstein,Corkum,Krausz,Ghimire,Bauer}. The mechanism of HHG can be specified by the three-step model \cite{Lewenstein,Corkum,Vampa1}. High-harmonic spectroscopy also paves an avenue to probe the ultrafast quasiparticle dynamics in condensed matters and to reconstruct Bloch-band structure, Berry curvature, electron correlation and shift vector of systems \cite{Schmid,Luu,Qian,Baykusheva,Huangkai}. High-order sideband generation (HSG), which is analog to HHG, has also attracted much attention over the past decade \cite{Liu1,Zaks,Crosse,Xie}. For HSG, the near-resonant excitation by a weak field is utilized to generate electron-hole pairs around the band edge, then the electron-hole pairs are driven by a strong long-wavelength light field, and when the driving field direction is reversed the electrons will recombine with holes to lead the radiations of high-order sideband photons \cite{Liu1,Zaks}. 

HSG can provide a way for exploring the topological properties of non-trivial states in condensed matters \cite{Yang,Wu1,Borsch,Banks}. In addition, compared with HHG, the electronic structure of Floquet matters can be delicately dressed by the laser parameters of the periodic infrared field. The HSG in experiments reflected dynamics in an open quantum system, in which quantum decoherence resulted in the extreme attenuation of its high-sideband signals \cite{Zaks} and further limited the HSG spectrum to appear as a single plateau. The aspect that how to extend the HSG spectral plateau is significant for probing the highly non-trivial quantum state. The electron-hole pairs are elementary excitations caused by infrared lasers. Previously, the process of electron-hole pair generation is depicted by the perturbative nonlinear optical framework  \cite{Liu1,Zaks}. In HSG, the non-perturbative effect in light-dressed states of  matters has not got enough attentions and has even been neglected. In this work, we revisit the significance of light-dressed states and find that these light-dressed states give rise to the multiple plateaus of the HSG spectrum. The plateau extension of HSG spectrum is also useful in many electro-optical applications such as wide-band optical multiplexers, optical pulses with ultra-high repetition rate, and optical communication with THz bandwidth.

This paper is organized as follows. In Sec. 2, we present the Floquet-engineering systems, Lindblad master equation and Semiconductor bloch equation. In Sec. 3, we systematically investigate the HSG spectra in Floquet matters. We summarize our paper in Sec. 4. Atomic units are used throughout this paper, unless specified otherwise.

\section{Theoretical method}
\subsection{Floquet-engineering systems}
\begin{figure}[t]
	\includegraphics[width=11 cm,height = 7 cm ]{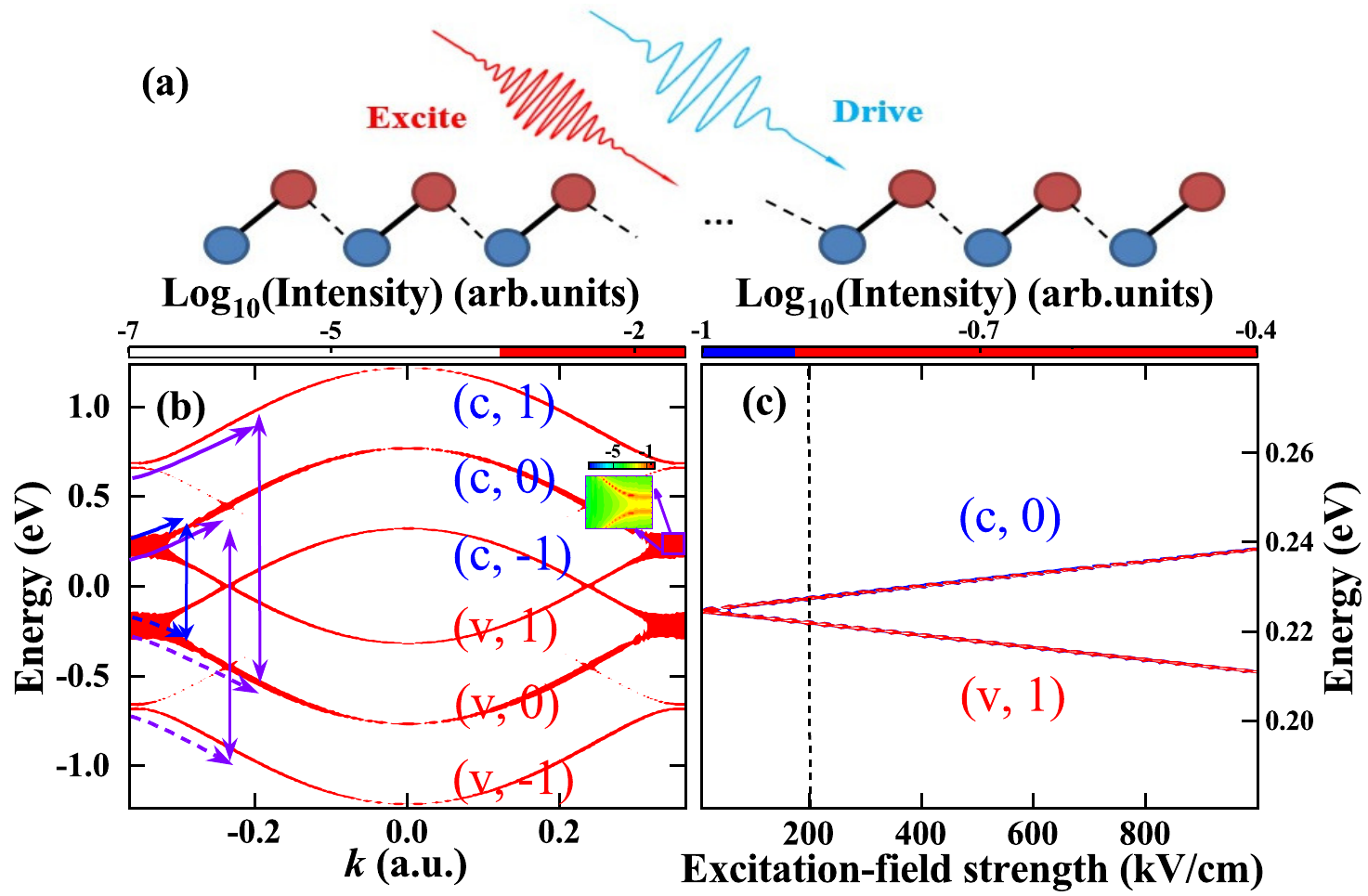} \centering
	\caption{(a) Schematic diagram of SSH chain under external field. (b) Floquet-Bloch bands (or spectral function) of the trivial bulk phase of SSH systems. $(\{\text{c},\text{v}\}, m)$ represent the field-dressed bands under number $m$ photons. The vertically double-headed arrows in (b) distinguish the electron-hole recombination between various Floquet-Bloch bands. A small panel in (b) highlights the characteristic gap opening. (c) The extent of the characteristic gap opening varies with the strength of the excitation field, with the vertical dashed line indicating the gap opening achieved under the adopted strength.
	} 	\label{Fig1}
\end{figure}

We commence by directing our attention to the Su-Schrieffer-Heeger (SSH) system, notable for the emergence of topological edge states \cite{Asb,Su,Li,Chen}.  Specifically, we examine linear chains comprising  500 lattice sites subject to open-boundary conditions. The application of such a system in HHG has been well studied \cite{JB,Ma}. In Fig. \ref{Fig1}(a) we give a schematic of the system. The system's Hamiltonian reads
\begin{equation}\label{Eq1}
	\hat{H}_0=\sum_{j}^{N} \left ( h_1 \hat{c}_{j,A}^{\dagger} \hat{c}_{j,B} + \text{H.c.} \right ) + \sum_{j}^{N-1} \left (h_2 \hat{c}_{j,B}^{\dagger} \hat{c}_{j+1,A}+\text{H.c}.\right ),
\end{equation}
where $\hat{c}_{j,A/B}^{\dagger}(\hat{c}_{j,A/B})$ is the generation (annihilation) operator in two sublattices A/B after the quadratic quantization. $h_1$ = -0.01827 a.u. and $h_2$ =  -0.01003  a.u. respectively represent intra- and inter-cellular hopping energies, and characterize two staggered hopping terms of the trivial phase. These parameters are applicable as models for quasi-one-dimensional systems, including conjugated polymers, organic crystals, carbon nanotubes, ferromagnetic perovskites, carbon chains, transition metal complexes, and organic charge transfer salts \cite{Roth,Kagoshima}. By diagonalizing the Hamiltonian in position representation, all eigenstates can be obtained, in which the in-gap edge state is absent in the trivial phase involving in our simulations. 

Since the SSH system has spatial translation periodicity, the Fourier transform of its coordinate-space Hamilton can be taken, i.e, $\hat{c}_{j, A}^{\dagger}=\frac{1}{\sqrt{N}} \sum_{k_0} \hat{c}^{\dagger}_{k_0, A} e^{i ak_0 j}$ and $\hat{c}_{j, B}^{\dagger}=\frac{1}{\sqrt{N}} \sum_{k_0}
\hat{c}^{\dagger}_{k_0, B} e^{i ak_0 j}$. $a$ is the primitive cell constant \cite{Su,JB,Asb}. Thus we obtain the momentum-space Hamiltonian $\hat{H}(k_0)=[h_1+h_2 \cos(ak_0)] \sigma_{x}+[h_2 \sin(ak_0)] \sigma_{y}$, in which $\sigma_{x,y}$ is the Pauli matrix. The sum of all $k_0$ will give the bulk Hamiltonian of the system. Given that the SSH system of bulk states is a two-band system, we straightforwardly designate the upper conduction band as the "c-band" and the lower valence band as the "v-band". Notably, the minimal bandgap for the bulk states is 0.448 eV, and the system is adopted to better construct the light-dressed Floquet-Bloch (FB) state \cite{Bao}.

The time-dependent expression of $\hat{H}(k)$ can be derived by substituting $k$ with $k(t) = k_{0} + A_{\text{ex}}(t)$. Note that the discussions of Floquet theory only involve the excitation field $A_{\text{ex}}(t)$. In this context, $k_{0}$ denotes the initial crystal momentum. Under the driving terahertz (THz) field, the time-dependent evolution of systems can be numerically achieved by solving the quantum master equation, which will be specified in the following subsection. It's worth noting that the external-field vector potential $A(t)$ here include both of the excitation $A_{\text{ex}}(t)$  and THz driving $A_{\text{dr}}(t)$ fields. For the real-space Hamiltonian, its time-dependent form is satisfied as ${\hat H(t)}$ = ${\text{e}^{-iA(t)\hat x}}{\hat H_0}{\text{e}^{iA(t)\hat x}}$ under the velocity gauge, in which $\hat x$ is the coordinate operator.

To reveal the role of non-perturbative effect in this light-dressed system, we turn to its Floquet Hamiltonian \cite{Flo_LRB}. According to Floquet's theory, a time-periodic system satisfies $\psi (t)= \varphi (t)e^{-i{\varepsilon}t }$, where $\varphi(t)$ represents the periodic component. Simultaneously, the time-periodic Hamiltonian ${\hat H(t)}$ can be expanded as the Fourier series of a fundamental frequency $\Omega$, where $\Omega$ is the light frequency of the coupling light field, thus constructing a Hilbert space and determining a series of FB states in a certain frequency range \cite{Floquet,Eckardt,Rudner3} (see Appendix A for more details). The following Floquet Hamiltonian can be constructed as
\begin{equation}\label{Eq2}
	\hat{\mathcal{H}}_\mathcal{F}=\left(\begin{array}{cccc}
		\ddots & \hat{H}_{-1} & \hat{H}_{-2} & \\
		\hat{H}_{1} & \hat{H}_{0}-m \hbar \Omega & \hat{H}_{(-1)} & \hat{H}_{-2} \\
		\hat{H}_{2} & \hat{H}_{1} & \hat{H}_{0}-(m+1) \hbar \Omega & \hat{H}_{-1} \\
		& \hat{H}_{2} & \hat{H}_{1} & \ddots
	\end{array}\right),
\end{equation}
where $\hat{\mathcal{H}}_\mathcal{F}$ satisfies $\hat{\mathcal{H}}_\mathcal{F}\varphi _n$ = $\varepsilon_n \varphi _n$ and $\varphi _n$ = $\begin{pmatrix}
	...&\phi^n_{m-1} &\phi^n_{m}  &\phi^n_{m+1}&...
\end{pmatrix}^T$. $\hat{H}_l$ is the Fourier expansion coefficient of $\hat{H}(t)$ = $\sum_{l}e^{il{\Omega}t}\hat{H}_l$, and $\phi_{m}$ is the $m$-th Fourier coefficient of $\varphi(t)$ \cite{Yan1,Yan2,Nagai,Jin,Neufeld,Ikeda,Faisal}. $\hat{\mathcal{H}}_\mathcal{F}$ represents an extended Hamiltonian, we truncated it and traversed the whole momentum space to determine the FB states we need.
As presented in Fig. \ref{Fig1}(b), we define the spectral function of the FB band structure as $S(k,E)$ = $\sum_{n}\delta(E-\varepsilon_n)\left|\langle k|\varphi_{n} \rangle \right |^2$ \cite{Huangkai}, in which $\left|\langle k|\varphi_{n} \rangle \right |^2$ = $\sum_j \langle \varphi_{j,n}| e^{-ikja_{0}} \sum_{j^{'}} e^{ikj'a_{0}}  |\varphi_{j', n} \rangle$. The system's eigenfunction $\left|\varphi_{n}\right\rangle$ with eigenvalue $\varepsilon_n$ is obtained from the diagonalization of Eq. (\ref{Eq2}), and $\left|\varphi_{j,n}\right\rangle$ is the value of eigenfunction $\left|\varphi_{n}\right\rangle$ at the lattice site $j$. $k$ is the quasimomentum. The Dirac function is broadened by using a Lorentzian shape $\frac{1}{\pi}\frac{\eta}{(E-\varepsilon_n)^2 + \eta^2}$ with a proper broadening factor $\eta$ = 0.001$|h_1|$.

The Floquet engineering system undergoes precise manipulation through the utilization of a near-resonant excitation field with frequency $\omega_{ex}$ ($\lambda \simeq $ 2700 nm), where the excitation field is considered to be a continuous wave. The strength of the excitation light pulses is fixed at 200 kV/cm, however its small vector potential makes it challenging to induce motion of electrons and holes. In Fig. \ref{Fig1}(b), the FB bands of the SSH systems could be obtained from the diagonalization of the Hamiltonian's matrix of Eq. (\ref{Eq2}) under the frequency space.

The index $(\{\text{c},\text{v}\}, m)$ is marked as the FB state with index $m$. Under such excitation field strength, it is sufficient to produce a non-perturbative modulation of the bulk states, and a characteristic gap opening emerges in their respective modes of FB bands as shown in Fig. \ref{Fig1}(b). Moreover, in experiments this infrared laser field with such light strength was also sufficient to produce the light-dressed FB states \cite{McIver,Kobayashi,Zhou}. In Fig. \ref{Fig1}(c), the variation of this light-dressed energy gap between bulk state (v, 1) and (c, 0) with the field strength of the excitation field is given. As the excitation field strength increases, the degree of energy gap opening gradually increases, convincingly exhibiting the characteristics of Floquet materials. The excitation field not only creates a Floquet system but also serves as a nearly resonant excitation field in the HSG process. Simultaneously, the driving THz field with the frequency $\omega$ (= 1 THz) shines the Floquet engineering systems. However, the magnitude of this gap in  Fig. \ref{Fig1}(b) is insufficient to significantly influence the overall resonant excitation process or the structure of the HSG plateaus.

\begin{figure}[t]
	\includegraphics[width=12 cm, height = 8 cm ]{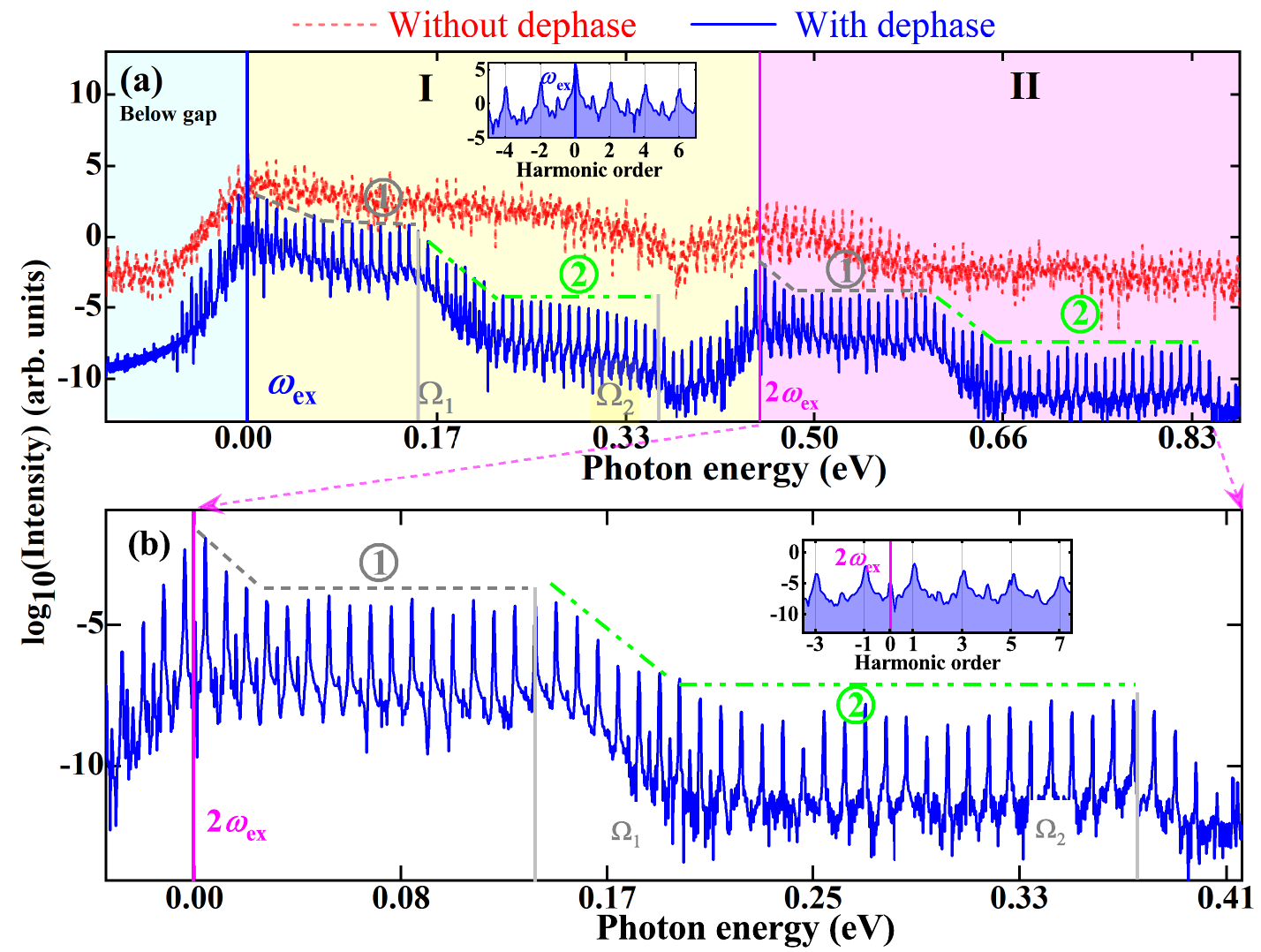}  \centering
	\caption{(a) Red dashed and blue solid curves respectively show the HSG spectra without and with decoherence, in which $T_2$ = 250 fs is considered. (b) Enlargement of the magenta spectral region II in (a). Inset panels in (a) and (b) give the odd-even order details of high-order sidebands. The zeros in (a) and (b) represent the reference point or baseline for the plotted HSG spectrum, in which the energies of the coupled photon number have been subtracted separately. The sidebands in the spectral I and II regions respectively present the even- and odd-order features. The symbols of \ding{172} and \ding{173} respectively denote two HSG plateaus in each spectral region.} 	\label{Fig2}
\end{figure}

\subsection{Lindblad master equation}
In our simulations, the directions of the linearly polarized THz and infrared lights are mutually parallel. In the time-dependent evolution process, we consider $A(t)$ = $A_{\text{ex}}(t)$ + $A_{\text{dr}}(t)$. $A(t)$ denotes the total vector potential including the excitation field $A_{\text{ex}}$ and driving field $A_{\text{dr}}$, where the THz driving field irradiates the Floquet systems tailored by infrared excitation light. The density matrix evolution of the system can be conducted either in position or momentum space. Initially, in position space, we have conducted the evolution under the framework of the Lindblad master equation. The time-dependent evolution of the density matrix satisfies $\frac{d \hat{\rho}(t)}{d t}$ = -$i[\hat{H}(t), \hat{\rho}]$ + $D(\hat{\rho})$. All states in valence band were adopted as the initial states in the trivial phase of SSH system. For the various interband relaxation mechanisms, we incorporate the pure dephasing term into the Von Neumann equation and indicate it as $\mathit{D}(\hat \rho)$ = $- \frac{{(1 - {\delta _{mn}})}}{{{T_2}}}{\hat \rho _{mn}}$. $T_2$ is set as 250 fs. The current operator is denoted as 
\begin{equation}\label{Eq3}
	\hat{j}=-i \sum_{m}\left(r_{m}-r_{m+1}\right) \hat{H}_{m, m+1} \hat{c}_{m}^{\dagger} \hat{c}_{m+1}-\text {H.c.},
\end{equation} 
where $r_{m}$ denotes the coordinate of the sublattice site. The light-induced current is written as $\mathbf{J}(t)$ = $\operatorname{Tr}\left(\hat{\rho}(t) \hat{j}\right)$. HSG spectrum can be obtained from the Fourier transform of the current. The emitted spectrum is characterized as the high-order sideband $2N\omega$ around the excitation light frequency $\omega_{\text{ex}}$.

\subsection{Semiconductor Bloch equation}
For the initial states, their evolution can also follow the formulation of semiconductor Bloch equations (SBEs) in momentum space \cite{Rana,Vampa2,Li2}. Time-dependent Hamiltonian in momentum space is written as
\begin{equation}\label{Eq4} 
	\hat{H}[k(t)] = 
	\begin{pmatrix}
		0  & h_1+h_2e^{ia[k(t)]}\\
		h_1+h_2e^{-ia[k(t)]}  & 0
	\end{pmatrix}.
\end{equation}
After diagonalization of $\hat{H}[k(t)]$, we have the instantaneous eigenstates and eigenvalues of the system. 
The time-dependent evolution of the density matrix $ \rho_{m n}^{{k}}$ can be obtained by the semiconductor Bloch equation under the Huston basis \cite{Mrudul,Golde,Vampa2}.
\begin{equation}\label{Eq5}
	\begin{split}
		\partial_{t} \rho_{m n}^{{k}}=  & -i \epsilon_{m n}^{{k_0}+{A}(t)} \rho_{m n}^{{k}}-\left(1-\delta_{m n}\right) \frac{\rho_{m n}^{{k}}}{T_{2}}	\\
		& +i {F}(t) \cdot\left[\sum_{l}\left({d}_{m l}^{{k_0}+{A}(t)} \rho_{l n}^{{k}}-{d}_{l n}^{{k_0}+{A}(t)} \rho_{m l}^{{k}}\right)\right].
	\end{split}
\end{equation}
Here $\epsilon_{m n}$ is the energy gap between band indexes $m$ and $n$. ${d}_{m n}$ is the dipole matrix elements calculated as $ {d}_{m n}=i\left \langle u_m  | \bigtriangledown_k|u_n  \right \rangle $, where $u_m$ and $u_n$ are the  eigenstates after diagonalization of the Hamiltonian of Eq. (\ref{Eq4}). $F(t)$ = $-\frac{dA(t)}{dt}$ is the electric field of laser. In addition, to improve the efficiency and convergence of the calculation, Eq. (\ref{Eq5}) is rewritten as the following form

	\begin{equation}\label{Eq6}
		\frac{\partial }{\partial t} \begin{pmatrix}
			n_{\text{cv}}^{{k}}\\
			\rho_{\text{cv}}^{{k}}\\
			\rho_{\text{vc}}^{{k}}
		\end{pmatrix}= \begin{pmatrix}
			0& -2iF(t){d}_{\text{cv}}({k})  & 2iF(t){d}_{\text{cv}}({k}) \\
			-iF(t){d}_{\text{cv}}({k}) & i\varepsilon _{\text{cv}}+ iF(t)[{d}_{\text{cc}}({k}) -{d}_{\text{vv}}({k}) ]-\frac{1}{T_2} & 0\\
			iF(t){d}_{\text{vc}}({k}) & 0 &i\varepsilon _{\text{vc}}+iF(t)[{d}_{\text{vv}}({k}) -{d}_{\text{cc}}({k}) ]-\frac{1}{T_2} 
		\end{pmatrix}\begin{pmatrix}
			n_{\text{cv}}^{{k}}\\
			\rho_{\text{cv}}^{{k}}\\
			\rho_{\text{vc}}^{{k}}
		\end{pmatrix}.
	\end{equation}

The evolution of Eq. (\ref{Eq6}) can be computed using the Crank-Nicolson method \cite{zhaoyp,dtyol,song}. $n_{\text{cv}}$ is the population difference between two bands and we set the $n_{\text{cv}}$ value at the initial moment to be $-1$. The resulting current can be expressed by the following term
\begin{equation}\label{Eq7}
	\begin{aligned}
		\mathbf{J}(t) & =\sum_{n} \int_{B Z} \rho_{n n}^{{k}} \nabla_{{k}} \varepsilon_{n}({k}) d^{3} k \\
		& +\sum_{n \neq m} \int_{B Z} i \varepsilon_{n m}({k}) \rho_{m n}^{{k}} {d}_{n m}({k}) d^{3} k .
	\end{aligned}
\end{equation}
In the right of Eq. (\ref{Eq7}), the first term is the intraband current, and the second term is the interband current. It is worth noting that SBEs requiring the periodic boundary condition only works in the momentum space and can not address the open-boundary systems with in-gap edge state.

\section{Discussion}

In Fig. \ref{Fig2}(a), we present the HSG spectra in the trivial phase of the SSH system, where the role of dephasing characterizes the multiple HSG plateaus clearer. The THz driving field strength is 38.5 kV/cm with a duration of sixteen optical cycles. For analytical convenience, we categorize the spectra into three distinct regions: the region below the gap, region I, and region II. While the observation of the two-plateau structure in spectral region I aligns with previous finding \cite{Crosse}, the emergence of spectral region II represents a novel discovery. In this spectral region II, the underlying mechanism pave an avenue to regulate HSG.

Considering the fact that the input optical field $\omega_{\text{ex}}$ creates the electron-hole pair and then the THz field drives the electron to recombine with the hole leading to the emissions of high-order sidebands, we shift the sideband orders by $\omega_{\text{ex}}$ in Fig. \ref{Fig2}(a). Due to the system's inversion symmetry the sideband orders of spectral region I satisfy the even orders $2N\omega$, which is exhibited by the inset panel in Fig. \ref{Fig2}(a). To unravel the spectral characteristics in region II, we delicately shift their sidebands by 2$\omega_{\text{ex}}$ and further display this spectral region in Figs. \ref{Fig2}(b). Contrary to the even-order sidebands in the spectral region I, the sidebands of region II are odd orders in Fig. \ref{Fig2}(b), in which this odd-order feature is robust under various detuning of excitation light. Its mechanism will be discussed subsequently. Then we turn to illustrate the cutoff frequencies in two spectral regions. One sees conventional HSG spectra with a two-plateau structure in both region I and region II, and further confirms their two cutoff frequencies $\Omega_1$ = 3.17$U_{p}$ and $\Omega_2$ = 8$U_{p}$, where $U_{p}$ =  ${A_0}^{2}/4m_{R}$ is the ponderomotive energy. $m_R$ (= 0.35 a.u.) is the electron-hole reduced mass of the bulk bands, and $A_0$ is the peak of the vector potential of the THz field. The formation of a two-plateau structure in spectral regions I and II is attributed to the quantum coherence among the recombination wave packets of electron-hole pair \cite{Du1,Crosse,WangGan}. Intriguingly, in Fig. \ref{Fig2}(a) region I extending delicately to region II holds a uniform spectral structure among them, which implies that this light-dressed system becomes the Floquet matter.

\begin{figure}[t]
	\includegraphics[width=12 cm, height = 4  cm ]{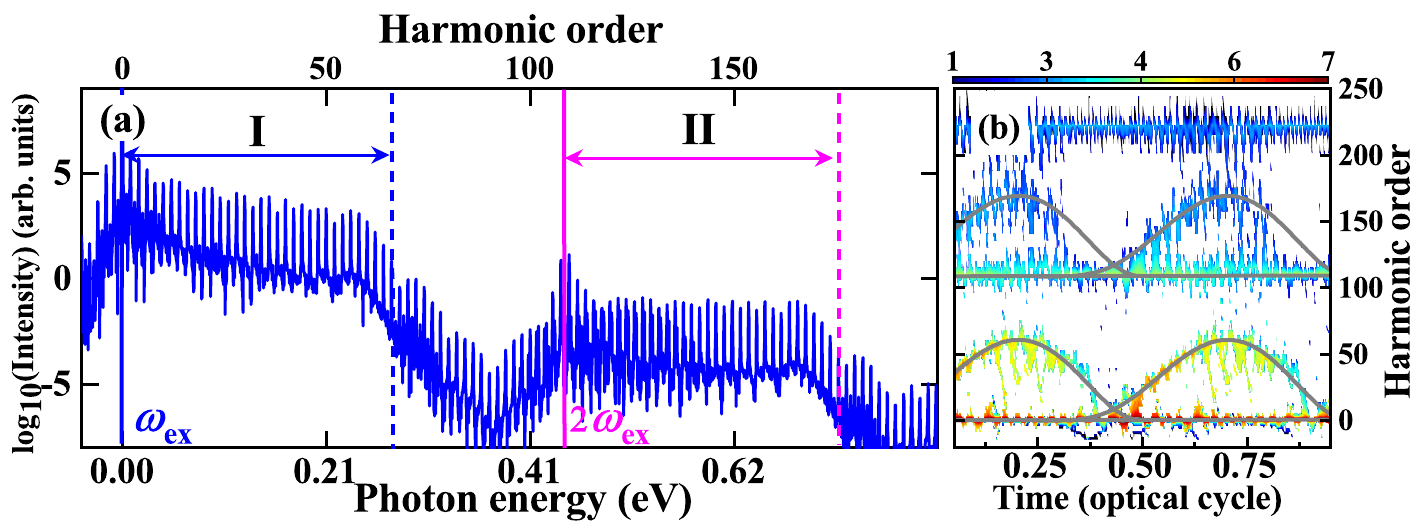}  \centering
	\caption{(a) HSG spectrum in Floquet matters. Under the case with a shorter dephasing time ($T_2$ = 150 fs), only the single plateau survives in each spectral region.  (b) Time-frequency analysis of the HSG spectra in (a), gray curves represent the predicted emission  timings from the semi-classical model.} 	\label{Fig3}
\end{figure}

Then, we will delve into the HSG of Floquet matters within the momentum space. To clarify the fact that the two-plateau structure in Fig. \ref{Fig2} attributes to the enough dephasing time, a shorter dephasing time ($T_{2}$ = 150 fs) has adopted in Fig. \ref{Fig3}. Here a stronger THz field with a strength of 51.5 kV/cm is involved in Fig. \ref{Fig3}(a). One could find that the two-plateau spectral structure in Fig. \ref{Fig2}(a) becomes the single spectral plateau in Fig. \ref{Fig3}(a). And in Fig. \ref{Fig3}(a) the odd-even characteristics of sidebands are robust. For the electron or hole in the FB band, its group velocity $v^{g}_{n} (n=e, h)$ can be written as $\upsilon_{n}^{g}(k)$ = $\frac{\partial \varepsilon_{n}(k)}{\partial k}|_{k}$. The relative displacement between electron and hole is denoted as $\Delta x$ =  $\int_{t^{'}}^{t} \{ v_{e}^{g}[k(\tau)] - v_{h}^{g}[k(\tau)] \} d \tau$. Zero displacement signifies the electron-hole recombination process occurring between the ionization timing $t^{\prime}$ and the recollision timing $t$, resulting in the emission of high-order sidebands with bandgap energy. This allows one to derive the trajectories of electron-hole recombination in a semi-classical prediction \cite{Vampa3}. In Fig. \ref{Fig3}(b) we also perform the time-frequency analysis on the HSG spectrum of Fig. \ref{Fig3}(a). The spectral regions I and II characterize the same temporal trajectories of the electron-hole recombination, which are also verified by the semi-classical prediction marked by gray curves. Furthermore, in Fig. \ref{Fig1}(b) we could distinguish the contributions between FB bands. One could confirm that the spectral regions I and II are respectively dominated by $\{(\text{c}, 0)$ - $(\text{v}, 0)\}$ and $\{(\text{c}, 1)$ - $(\text{v}, 0)$ or $(\text{c}, 0)$ - $(\text{v}, -1)\}$ interband recombination of electron-hole pair. We can generalize the above rule to $\{(\text{c}, m)$ - $(\text{v}, n)\}$, when the index ($m - n$) = 1, it equivalently contributes to the spectral region II. Higher sideband plateaus are determined by electron-hole recollisions between FB bands with larger index of ($m - n$).

To elucidate the odd-even sideband properties in the spectroscopic regions I and II, we will discuss the dynamic symmetry between the FB states. One can also refer to the dynamical symmetry theory for further insights \cite{Neufeld}. For the SSH systems in momentum space, its Hamiltonian satisfies the inversion symmetry, i.e., $\hat{H}_0(-k)=\hat{H}_0^{\ast } (k)$. Keeping the external field with temporal symmetry $A(t+T/2)$ = $-A(t)$ in mind, the time-dependent Hamiltonian obeys $\hat{H}(-k,t+T/2)$ = $\hat{H}^{\ast } (k,t)$, where $T=2\pi/\Omega$ is the optical period of laser fields. Therefore, for the time-dependent Hamiltonian $\hat{H}(\pm k,t)$, its Fourier series in temporal expansion are denoted as $ \hat{H}(t)$ = $\sum_{l}e^{il{\Omega}t}\hat{H}_l$, in which the coefficient satisfies $\hat{H}_l(-k)$ = $(-1)^{l}\hat{H}_l^\ast(k)$ with integer $l$. 
Based on the mentioned symmetry differences between $\hat{H}_l({\pm}k)$, one can derive a rule as $\hat{H}_{m n}^{\mathrm{F}}(-k)=(-1)^{m-n} \hat{H}_{m n}^{\mathrm{F}\ {\ast }}(k)$, in which $\hat{H}_{mn}^F$ is the matrix element $\hat{H}_{m-n}$ in Eq. (\ref{Eq2}) and satisfies $\hat{H}_{mn}^F(k)\phi^\alpha _n(k)=\varepsilon^\alpha(k) \phi^\alpha _m(k)$. $m$ or $n$ represents the index of FB state. Under the appropriate selected overall phase, the wave function will obey $\phi _n(-k)$ = $(-1)^{n}\phi _n^{\ast }(k)$, where $n$ is the index of FB state \cite{Ikeda}. Thus, the symmetry characteristics encoded in the coupling FB wave functions can determine the order feature of the sideband.

\begin{figure}[t]
	\includegraphics[width= 12 cm, height = 6 cm ]{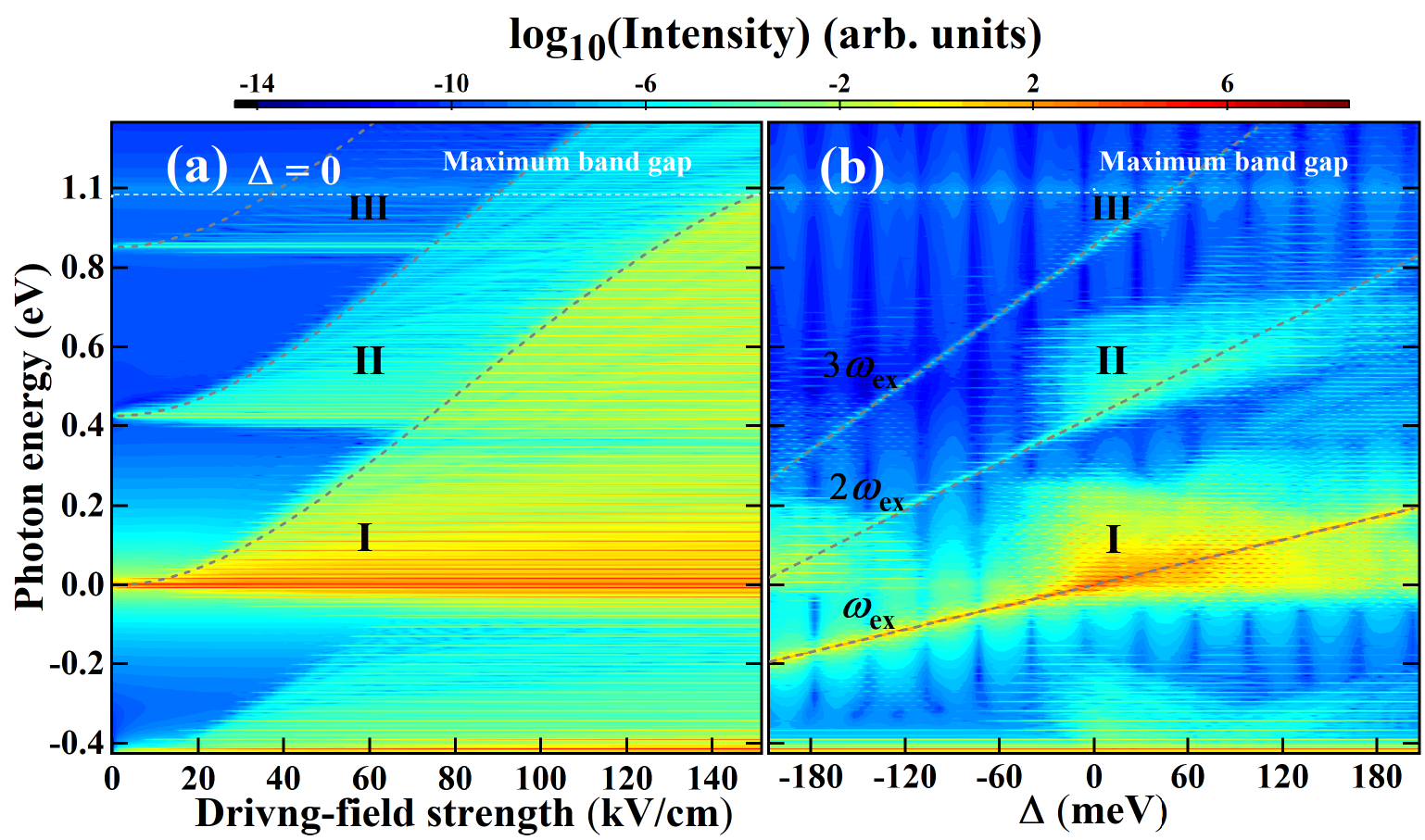} \centering
	\caption{(a) HSG spectra varies with the driving-field strength. Cutoff energy of three regions marked by the gray dashed lines is predicted by the semi-classical model. (b) HSG spectra as a function of the excitation-field frequency ($\Delta$ = $\omega_{\text{ex}}$ - $\epsilon_{g}$). Three gray dashed lines in (b) denote resonance peaks after coupled photons. Dephasing time is the same as that in Fig. \ref{Fig2}.} 	\label{Fig4}
\end{figure}

For the high-order sideband spectrum contributed by electron-hole recombination between band indexes $\{(\text{c}, m)$ $-$ $(\text{v}, n)\}$, its odd-even properties of sidebands are determined by the odd or even characteristic of the difference of the two FB index ($m - n$). When the index ($m - n$) is odd, the symmetries involved in two coupling FB states are opposite, leading to the final current satisfies: $J(k,N)=(-1)^{(N+2)}J(-k,N)$, where N represents the frequency component of current. Obviously, under the odd number $N$, the $\pm{k}$ channels cancel each other out so that only the even-order frequency component survives leading to odd-order sidebands \cite{Jin,Neufeld,Ikeda,Faisal,Ma}. Conversely, if the FB band index ($m - n$) is even, the final current meets $J(k,N)=(-1)^{(N+1)}J(-k,N)$, we can deduce the even-order sidebands (See details in Appendix B).

Further, we investigate HSG in Floquet materials by varying the laser-field parameters.
In Figs. \ref{Fig4}(a) and  \ref{Fig4}(b), we respectively unravel the roles of the driving-light  field strength and the excitation-light detuning ($\Delta$ = $\omega_{\text{ex}}$ - $\epsilon_{g}$) on HSG spectra. Here, $\epsilon_{g}$ is the minimum band gap between bulk bands. As depicted in Fig. \ref{Fig4}(a), the cutoff frequencies in spectral regions I, II, and III adhere to the scaling rule of 3.17$U_p$ predicted by the semi-classical model.
It is evident from Fig. \ref{Fig4}(a) that the FB modes dominate their respective HSG spectral regions. In conventional HHG, the broadening of the frequency comb is typically limited by the maximal gap of the energy band structure \cite{Bielke}. However, such limitations can be surpassed in Floquet-engineering systems, enabling the attainment of broader high-order sideband plateaus or continuous spectra. Additionally, in Fig. \ref{Fig4}(b), we observe the FB modes depicted as their respective slope lines, suggesting that Floquet materials can be finely tailored by the excitation light field, thereby providing a means to control HSG through Floquet engineering.

Moreover, in Fig. \ref{Fig4}(b) we observe that the high-order sideband plateau strongly depends on the positive laser detuning $\Delta$. When we tune the excitation light frequency to deviate from the situation of zero detuning, in spectral regions I, the spectral plateau width and the cutoff energies are shrunk from 3.17$U_p$ under a positive laser detuning. In spectral region II, it is noteworthy that the energy gap between the ($\text{c}, m$) and ($\text{v}, m-1$) FB bands increases with the rising excitation-field frequency, leading to an extension of the cutoff energy in region II to a certain extent. Moreover, it is observed that the spectral width of region II decreases with increasing detuning. The narrowing of the spectral plateaus can be attributed to the fact that the electron created under the overexcitation of light possesses an initial velocity, which subsequently hinders its recombination with the hole. \cite{Xie}. 
Obviously, in the cases with negative detuning, HSG signal can be emitted when the excitation-field photon energy is near half of the minimal band gap (corresponding to $\Delta \simeq -210$ meV), which leads to a two-photon excitation process. The FB band provides a step for the generation of electron-hole pairs between the original valence and conduction bands of the field-free system. This two-photon excitation results in its high-sideband emissions around the 2$\omega_{\text{ex}}$ energy zone, as shown in the left side of Fig. \ref{Fig4}(b).

\section{Conclusion}
In summary, periodic excitation light generates Floquet matters under relatively weak intensity, and a unique HSG in Floquet matters has been revealed. The multiple-plateau HSG exhibits unprecedented features beyond the single-plateau HSG spectra. Furthermore, the electronic structure of Floquet matters can be regulated by the periodic excitation light, offering a potent method for controlling the HSG process, although it should be noted that this Floquet system has a relatively low damage threshold. The Floquet-Bloch and dynamical-symmetry theories can clarify this multiple-plateau HSG spectrum and its odd-even characteristics. This work demonstrates how Floquet engineering offers a new avenue for controlling HSG, providing a promising approach to explore non-equilibrium quantum states within Floquet matter. Beyond HHG and HSG studies, light-dressed Floquet matter is an issue of interest and an extremely active and exciting field. Our work also shows a direct time-domain insight into Floquet physics and explores the fundamental frontiers of ultrafast band-structure engineering.

\section*{APPENDIX A: Floquet theory}
\setcounter{equation}{0}
\renewcommand{\theequation}{A\arabic{equation}}
Consider that most spectroscopic methods satisfy a fundamental property, i.e., under the action of periodic external fields. For a periodically driven field with frequency $\Omega={2\pi/T}$, where $T$ is an optical period. The wave function satisfies periodicity, and the solution can be written in the following form \cite{Floquet,Eckardt,Rudner3}
\begin{equation}\label{A1}
	\psi (t)= \varphi (t)e^{-i{\varepsilon}t },
\end{equation}
where $\varphi(t)=\varphi(t+T)$. This is the expression of Floquet theory which is similar to Bloch theorem. For a solution that satisfies the time periodicity, we use the discrete Fourier transformation to determine the frequency distribution, that is $\varphi(t)= \sum_{n}e^{in{\Omega}t}\phi_n$. Then one substitutes Eq. (\ref{Eq4}) to the Schr\"odinger equation and to get the following solution 
\begin{equation}\label{A2}
	\left [\hat{H}(t)-i\hbar \frac{\partial }{\partial t}  \right ]\varphi(t)=\varepsilon\varphi(t).
\end{equation}
$\hat{H}(t)$ is the time-dependent Hamiltonian and can be expanded as the discrete Fourier series, i.e.,
\begin{equation}\label{A3}
	\hat{H}(t)=\sum_{l}e^{il{\Omega}t}\hat{H}_l.
\end{equation} 
We substitutes $\varphi(t)= \sum_{n}e^{in{\Omega}t}\phi_n$ to Eq. (\ref{A2}), and set $\hat{\mathcal{H}}_\mathcal{F}=\hat{H}(t)-n\hbar\Omega$, which satisfies $\hat{\mathcal{H}}_\mathcal{F}\phi_n=\varepsilon\phi_n$. Thus construct an extended Hilbert space $\left | \alpha,n   \right \rangle$=$\left | \alpha  \right \rangle \otimes \left | n  \right \rangle$=$\left | \alpha  \right \rangle{e^{in{\Omega}t} }$. $\left|{\alpha}\right \rangle$ is the internal degree of freedom. In the momentum space of the two-band model, this degree of freedom is two. $\left |{n}  \right \rangle$ is the degree of freedom that couples with the external-field frequency. The matrix element of Floquet Hamiltonian $\hat{\mathcal{H}}_\mathcal{F}$ is denoted as
\begin{equation}\label{A4}
	\begin{aligned}
		\hat{H}_{mn}^F&=\left\langle\alpha^{\prime}, m\left|\hat{\mathcal{H}}_\mathcal{F}\right| \alpha, n\right\rangle\\
		&=\left\langle\alpha^{\prime}\left|\sum_{l}\hat{H}_{l} e^{i l{ \Omega} t} e^{-i m\Omega t} e^{i n \Omega t}\right| \alpha\right\rangle -n \hbar \Omega\left\langle\alpha^{\prime}, m \mid \alpha, n\right\rangle \\
		&=\left\langle\alpha^{\prime}\left|\hat{H}_{m-n}\right| \alpha\right\rangle-n \hbar \Omega \delta _{\alpha \alpha^{\prime}}\delta_{n m}.
	\end{aligned}
\end{equation}
Then we can construct a Hamiltonian in Floquet space. After traversing the whole momentum space, the desired Floquet-Bloch state can be obtained. 

\section*{APPENDIX B: symmetry and order characteristic analysis}
\setcounter{equation}{0}
\renewcommand{\theequation}{B\arabic{equation}}
To begin, we notice the following inversion symmetry of the SSH system \cite{Asb}:
\begin{equation}\label{B1}
	\hat{H}_0(k)=\hat{H}_0^{\ast } (-k)
\end{equation}
which implies the time-reversal symmetry of the SSH system. Since the external field follows $A(t+T_{ex}/2)$ = $-A(t)$ in the time domain, $T_{ex}=2{\pi}/{\Omega}$ is the excitation field optical period, the time-dependent Hamiltonian obeys $\hat{H}(k,t)$ = $\hat{H}^{\ast } (-k,t+T_{ex}/2)$.
We give the following proof:
\begin{subequations}\label{B2} \small
	\begin{equation}\label{B2a}
		\hat{H}(k,t)=\sum_{n}e^{in{\Omega}t}\hat{H}_{n}(k),\\
	\end{equation}
	\begin{equation}\label{B2b}
		\hat{H}^{\ast}(-k,t+T_{ex}/2)=\sum_{n}e^{in{\Omega}(t+T_{ex}/2)}\hat{H}^{\ast}_{n}(-k),\\
	\end{equation}
	\begin{equation}\label{B2c}
		\hat{H}_n(k)=e^{in{\pi}}\hat{H}^{\ast}_n(-k).
	\end{equation}
\end{subequations}
The left of the equal sign of (\ref{B2a}) and (\ref{B2b}) are equal, thus (\ref{B2c}) is easy to get. For the SSH system, it satisfies $\hat{H}_{n}(k)$ = $(-1)^n\hat{H}_{n}^{\ast }(-k)$, and thus the corresponding wave function also formulates $\phi _n(-k)$ = $(-1)^{n}\phi _n^{\ast }(k)$. Here n denotes the index of Floquet-Bloch state. This reveals the temporal inversion symmetry of the Floquet-Bloch states.

\begin{figure}[t]
	\includegraphics[width=11 cm,height = 5 cm ]{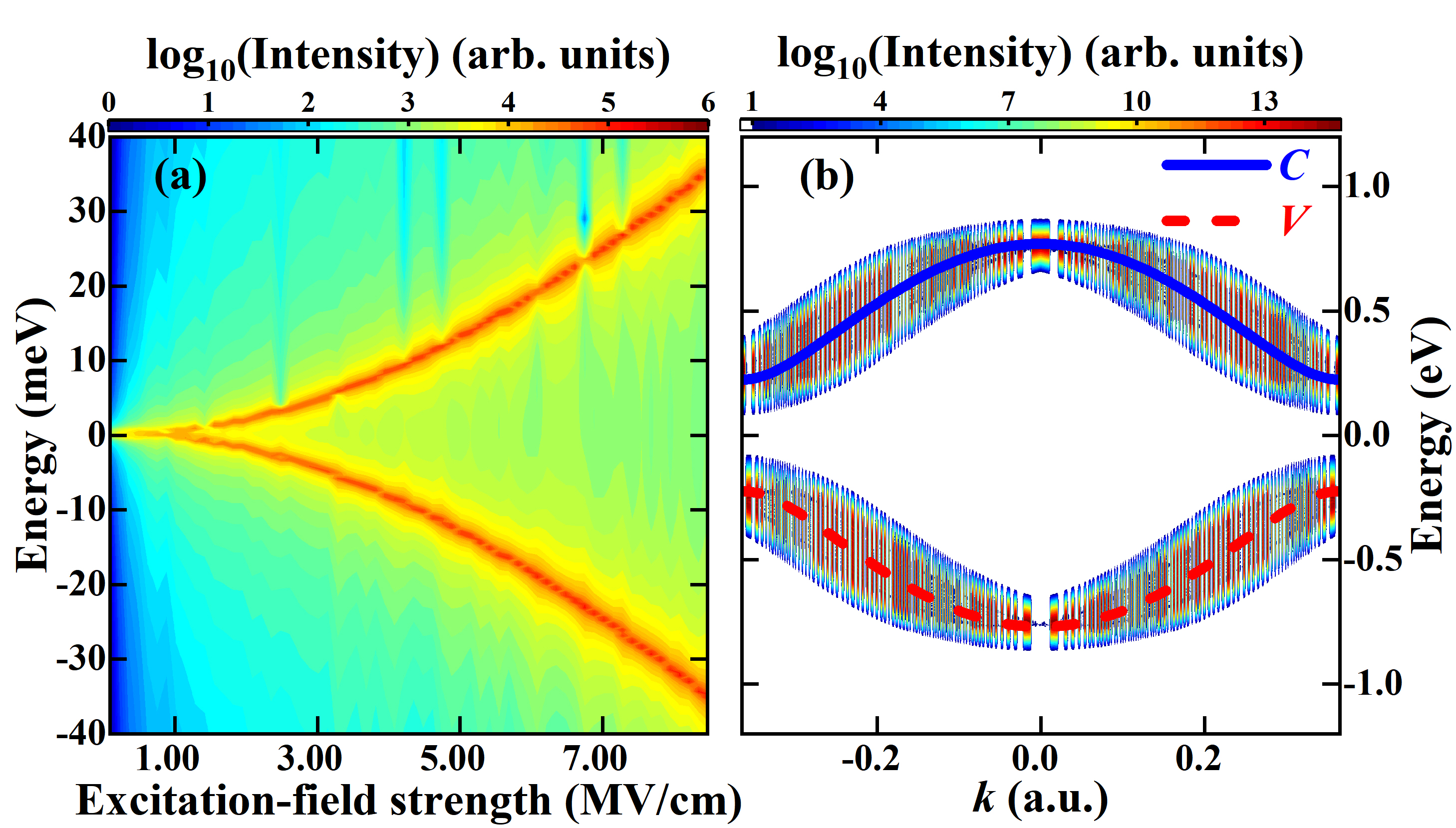}  \centering
	\caption{(a) The degree of Floquet-Bloch band opening varies with the excitation-field strength.  $k$ ($\simeq$ -0.235 a.u.) is set at band-crossing point between Floquet band $(\text{c}, -1)$ and $(\text{v}, 1)$. $(\{\text{c},\text{v}\}, m)$ represent valence and conduction bands coupling with number $m$ photon. (b) The Floquet-Bloch bands under the action of the THz field with laser strength 50 kV/cm. The field-free bands are denoted by blue and red curves in (b).} 	\label{SMFIG1}
\end{figure}

We then perform Floqeut theory analysis on the HSG emission \cite{Faisal,Ikeda}. Taking the Floquet-Bloch bands,  (c, 1) and (v, 0), as representative examples, their respective wave functions follow the rules  ${\phi^{\text{c}}}(k)$ = $-{\phi^{\text{c}}}^{\ast }(-k)$ and ${\phi^{\text{v}}}(k)$ = ${\phi^{\text{v}}} ^{\ast }(-k)$. Under the driving of the THz field with optical period $T_{\text{THz}}$ = $2{\pi}/{\omega}$, one can also exploit the Floquet theorem to expand their wave functions by Fourier series and to analyze the odd-even feature of high-order sidebands, which are denoted as
\begin{subequations}\label{B3}
	\begin{equation}
		\phi_{\text{c}}(k,t)=e^{i\varepsilon_{\text{c}}t}u^{c}(t)=e^{i\varepsilon_{\text{c}}t}\sum_{m}\chi_{m}^{\text{c}}(k)e^{im{\omega}t},
	\end{equation}
	\begin{equation}
		\phi_{\text{v}}(k,t)=e^{i\varepsilon_{\text{v}}t}u^{v}(t)=e^{i\varepsilon_{\text{v}}t}\sum_{n}\chi_{n}^{\text{v}}(k)e^{in{\omega}t}.
	\end{equation}
\end{subequations}
The time-dependent wave function satisfies the conditions: ${u^{\text{c}}}(k,t)$ = $-{u^{\text{c}}}^{\ast }(-k,t+T_{\text{THz}}/2)$ and ${u^{\text{v}}}(k,t)$ = ${u^{\text{v}}} ^{\ast }(-k,t+T_{\text{THz}}/2))$. According to the proof of Eq. (\ref{B2}), the conclusions can also be drawn as
\begin{subequations}\label{B4}   \small
	\begin{equation}
		\chi _m^{\text{c}}(k)=(-1)^{m+1}{\chi _m^{\text{c}}}^{\ast }(-k),
	\end{equation}
	\begin{equation}
		\chi _n^{\text{v}}(k)=(-1)^{n}{\chi _n^{\text{v}}}^{\ast }(-k).
	\end{equation}
\end{subequations}
To obtain the light-inducing current and extract the frequency components underlying the extremely nonlinear current, we define the current operator and perform Fourier series expansion, i.e., $\hat{j}(k,t)=\sum_{l}\hat{j}_{l}(k)e^{il{\omega}t}$, here $l$ is an integer. Since the current operators in ${\pm}k$ are opposite, leading the coefficient term $\hat{j}_{l}(k)$ possesses the following relationship as
\begin{equation}\label{B5}
	\hat{j}_{l}(k)=(-1)^{l+1}\hat{j}_{l}(-k).
\end{equation}
Its proof process is similar to Eq. (\ref{B2}). Then, we deduce the nonlinear current as
\begin{equation}\label{B6} \small 
	\begin{aligned} 
		J(t)&=\sum_{k} \left \langle \phi(k,t) |\hat{j}(k,t)| \phi(k,t)    \right \rangle \\
		&=\sum_{k}\sum_{{\alpha},{\beta }=\{\text{c},\text{v}\}}\sum_{m,n,l}\chi _{m} ^{\alpha}(k)^{\ast}\hat{j}_{l}(k)\chi _{n} ^{\beta}(k)\times e^{-i[{\varepsilon_{\alpha}(k)-{\varepsilon_{\beta}(k)}}+(m-l-n){\omega}]t}.\\
	\end{aligned}
\end{equation}
HSG is determined by the electron-hole recollision, we thus concern the interband recombination between two bands (${\alpha}, {\beta }=\text{c}, \text{v}$), in which the energy gap, $\varepsilon_{\alpha}(k)-{\varepsilon_{\beta}(k)}$, also obeys the time-reversal symmetry
of their energy bands. We perform a substitution $m=n+l+N$ in term $e^{-i(m-l-n){\omega}t}$, which could embody the frequency component $N$.

Substituting Eqs. (\ref{B4}) and (\ref{B5}) into Eq. (\ref{B6}), one obtains
\begin{subequations}\label{B7} \small
	\begin{equation}
		J(k,N)={\chi _{n+l+N}^{\text{c} }(k)}^{\ast }\hat{j}_{l}(k){\chi _{n}^{\text{v}}(k)},\\
	\end{equation}
	\begin{equation}
		J(-k,N)=(-1)^{(n+l+N)+1}(-1)^{(l+1)}(-1)^{n}{\chi _{n+l+N}^{\text{c} }(k)}^{\ast }\hat{j}_{l}(k){\chi _{n}^{\text{v}}(k)},\\
	\end{equation}
	\begin{equation}
		J(k,N)=(-1)^{(N+2)}J(-k,N).
	\end{equation}
\end{subequations}
For odd $N$, the channels of ${\pm}k$ cancel each other out, so that only even order signals eventually survive. Considering the HSG signal after resonance excitation, only odd-order sideband signals exist in the corresponding plateau. Extending the above rule to the electron-hole recombination signal between $\{(\text{c}, m)$ - $(\text{v}, n)\}$, it is obvious that when ($m$ - $n$) is odd, the corresponding HSG plateau signal conforms to the odd-order characteristics. Similarly, it can be given that for  even ($m$ - $n$), its current satisfies $J(k,N)=(-1)^{(N+1)}J(-k,N).$ The property of the corresponding plateau satisfies even order. This indicates that plateau regions exhibit either odd or even-order characteristics based on the differences of contributions from ($m$ - $n$).
\section*{APPENDIX C: Band inversion and Floquet-Bloch state under THz field}

Since the different Floquet states of the valence band and conduction band will form a cross, the study of such a band intersection is also desirable. As shown in Fig. \ref{SMFIG1}(a), the Floquet band (c, -1) and (v, 1) will cross each other. However, with the increasing of excitation field strength, the band inversion is realized to avoid band crossing. This result is consistent with the experimental results \cite{McIver,Zhou}. The signature of band inversion is indicative of a topological phase transition taking place within a non-equilibrium state. Future explorations could involve leveraging all-optical detection methods to delve deeper into these non-equilibrium processes. In addition, we show the Floquet-Bloch states under the individual action of a THz field with strength 50 kV/cm as presented in Fig. \ref{SMFIG1}(b). The THz field only performs the periodic energy modulation on the original energy bands, and it is difficult to achieve the non-trivial band inversion due to its small photon energy. Thus, the THz fields primarily play the role of driving the motion of electron-hole pairs.

\section*{Funding} T.-Y. Du is supported by the Natural Science Foundation of Hubei Province China (Grant No. 2023AFB613) and the Fundamental Research Funds for the Central Universities, China University of Geosciences (Wuhan) with Grant No. G1323523064; M. Li acknowledges the financial support from the Research Grant Council of Hong Kong (Project No.PolyU 25301522 and PolyU 15301323), National Natural Science Foundation of China (22373081).

\section*{Acknowledgement} We thank Prof. Ren-Bao Liu for the fruitful discussions. 

\section*{Disclosures} The authors declare no conflicts of interest.
\section*{Data Availability} All the data used in this study are available upon reasonable request from the corresponding author.




\begin{thebibliography}{99}
	\bibitem{Oka} T. Oka and H. Aoki, "Floquet theory of photo-induced topological phase transitions: application to graphene," Phys. Rev. B {\bf 79}(16), 081406(R) (2009).
	\bibitem{Linder} N. H. Lindner, G. Refael, and V. Galitski, "Floquet topological insulator in semiconductor quantum wells," Nat. Phys. {\bf 7}(6), 490 (2011).
	\bibitem{Torre} A. de la Torre, D. M. Kennes, M. Claassen, S. Gerber, J. W. McIver, and M. A. Sentef, "Colloquium: Nonthermal pathways to ultrafast control in quantum materials," Rev. Mod. Phys. {\bf 93}(4), 041002 (2021).
	\bibitem{Rudner1} M. S. Rudner and N. H. Lindner, "Band structure engineering and non-equilibrium dynamics in Floquet topological insulators," Nat. Rev. Phys. {\bf 2}(5), 229 (2020).
	\bibitem{Rechtsman} M. C. Rechtsman, J. M. Zeuner, Y. Plotnik, Y. Lumer, D. Podolsky, F. Dreisow, S. Nolte, M. Segev, and A. Szameit, "Photonic Floquet topological insulators," Nature (London) {\bf 496}(7444), 196 (2013).
	\bibitem{Minguzzi} J. Minguzzi, Z. Zhu, K. Sandholzer, A.-S. Walter, K. Viebahn, and T. Esslinger, "Topological pumping in a floquet-bloch band," Phys. Rev. Lett. {\bf 129}(5), 053201 (2022).
	\bibitem{Rudner2} M. S. Rudner, N. H. Lindner, E. Berg, and M. Levin, "Anomalous edge states and the bulk-edge correspondence for periodically driven two-dimensional systems," Phys. Rev. X {\bf 3}(3), 031005 (2013).
	\bibitem{Machado} B. Ye, F. Machado, and N. Y. Yao, "Floquet phases of matter via classical prethermalization," Phys. Rev. Lett. {\bf 127}(14), 140603 (2021).
	
	
	\bibitem{McIver}  J. W. McIver, B. Schulte, F.-U. Stein, T. Matsuyama, G. Jotzu, G. Meier, and A. Cavalleri, "Light-induced anomalous Hall effect in graphene," Nat. Phys. {\bf 16}(1), 38 (2020).
	\bibitem{Kobayashi} Y. Kobayashi, C. Heide, A. C. Johnson, V. Tiwari, F. Liu, D. A. Reis, T. F. Heinz, and S. Ghimire, "Floquet engineering of strongly driven excitons in monolayer tungsten disulfide," Nat Phys {\bf 19}(2), 171 (2023).
	\bibitem{Zhou} S. Zhou, C. Bao, B. Fan, H. Zhou, Q. Gao, H. Zhong, T. Lin, H. Liu, P. Yu, P. Tang \emph{et al}., Pseudospin-selective Floquet band engineering in black phosphorus," Nature (London) {\bf 614}(7946), 75 (2023).
	\bibitem{Wang} Y. H. Wang, H. Steinberg, P. Jarillo-Herrero, and N. Gedik, "Observation of Floquet-Bloch states on the surface of a topological insulator," Science {\bf 342}(6157), 453 (2013).
	\bibitem{Ito} S. Ito, M. Sch\"uler, M. Meierhofer, S. Schlauderer, J. Freudenstein, J. Reimann, D. Afanasiev, K. A. Kokh, O. E. Tereshchenko, J. G\"udde, M. A. Sentef, U. H\"ofer, and R. Huber, "Build-up and dephasing of Floquet–Bloch bands on subcycle timescales," Nature (London) {\bf 616}, 696 (2023).
	
	
	\bibitem{Lewenstein} M. Lewenstein, P. Balcou, M. Y. Ivanov, A. L’Huillier, and P. B. Corkum, "Theory of high-harmonic generation by low-frequency laser fields," Phys. Rev. A {\bf 49}(3), 2117 (1994).
	\bibitem{Corkum} P. B. Corkum, "Plasma perspective on strong field multiphoton ionization," Phys. Rev. Lett. {\bf 71}(13), 1994 (1993).
	\bibitem{Krausz} F. Krausz and M. Ivanov, "Attosecond physics," Rev. Mod. Phys. {\bf 81}(1), 163 (2009).
	\bibitem{Ghimire} S. Ghimire, A. D. DiChiara, E. Sistrunk, P. Agostini, L. F. DiMauro, and D. A. Reis, "Observation of high-order harmonic generation in a bulk crystal," Nat. Phys. {\bf7}(2), 138 (2011).
	\bibitem{Bauer} D. Bauer and K. K. Hansen,  "High-harmonic generation in solids with and without topological edge states," Phys. Rev. Lett. {\bf 120}(17), 177401 (2018).
	\bibitem{Vampa1} G. Vampa, T. J. Hammond, N. Thir\'e, B. E. Schmidt, F. L\'egar\'e, C. R. McDonald, T. Brabec, and P. B. Corkum, "Linking high harmonics from gases and solids," Nature (London) {\bf 522}(7557), 462 (2015).
	\bibitem{Schmid} C. P. Schmid, L. Weigl, P. Gr\"ossing, V. Junk, C. Gorini, S. Schlauderer, S. Ito, M. Meierhofer, N. Hofmann, D. Afanasiev, J. Crewse, K. A. Kokh, O. E. Tereshchenko, J. G\"udde, F. Evers, J. Wilhelm, K. Richter, U. H\"ofer, and R. Huber, "Tunable non-integer high-harmonic generation in a topological insulator," Nature (London) {\bf 593}(7859), 385 (2021).
	\bibitem{Luu} T. T. Luu, M. Garg, S. Y. Kruchinin, A. Moulet, M. T. Hassan, and E. Goulielmakis, "Extreme ultraviolet high-harmonic spectroscopy of solids," Nature {\bf 521}(7553), 498 (2015).
	\bibitem{Qian} C. Qian, C. Yu, S. C. Jiang, T. Zhang, J. C. Gao, S. Shi, H. Q. Pi, H. M. Weng, and R. Lu, "Role of shift vector in high-harmonic generation from noncentrosymmetric topological insulators under strong laser fields," Phys. Rev. X {\bf 12}(2), 021030 (2022).
	\bibitem{Baykusheva} D. Baykusheva, A. Chac\'on, D. Kim, D. E. Kim, D. A. Reis, and S. Ghimire, "Strong-field physics in three-dimensional topological insulators," Phys. Rev. A {\bf 103}(2), 023101 (2021).
	\bibitem{Huangkai} K. Huang and T.-Y. Du, "Quasiparticle picture for high-harmonic generation in correlated electron systems," Phys. Rev. B {\bf 108}(15), 155125 (2023).

	\bibitem{Liu1} R.-B. Liu and B.-F. Zhu, in \textit{Physics of Semiconductors:28th International Conference on the Physics of Semiconductors-ICPS 2006}, AIP Conf. Proc. No. 893 (AIP, New York, 2007), p. 1455.
	\bibitem{Zaks} B. Zaks, R.-B. Liu, and M. S. Sherwin, "Experimental observation of electron–hole recollisions," Nature {\bf 483}(7391), 580 (2012).
	\bibitem{Crosse} J. A. Crosse and R.-B. Liu, "Quantum-coherence-induced second plateau in high-sideband generation,"  Phys. Rev. B {\bf 89}(12), 121202(R) (2014).
	\bibitem{Xie} X. T. Xie, B. F. Zhu, and R. B. Liu, "Effects of excitation frequency on high-order terahertz sideband generation in semiconductors," New J. Phys. {\bf 15}(10), 105015 (2013).
	\bibitem{Wu1} Q. Wu and M. S. Sherwin, "Explicit formula for high-order sideband polarization by extreme tailoring of Feynman path integrals," Phys. Rev. B {\bf 107}(17), 174308 (2023).
	\bibitem{Yang} F. Yang, X. Xu, and R.-B. Liu, "Giant Faraday rotation induced by the Berry phase in bilayer graphene under strong terahertz fields," New J. Phys. {\bf 16}(4), 043014 (2014).
	\bibitem{Borsch} M. Borsch, C. P. Schmid, L. Weigl, S. Schlauderer, N. Hofmann, C. Lange, J. T. Steiner, S. W. Koch, R. Huber, and M. Kira, "Super-resolution lightwave tomography of electronic bands in quantum materials," Science {\bf 370}(6521), 1204 (2020).
	\bibitem{Banks} H. B. Banks, Q. Wu, D. C. Valovcin, S. Mack, A. C. Gossard, L. Pfeiffer, R.-B. Liu, and M. S. Sherwin, "Dynamical birefringence: electron-hole recollisions as probes of Berry curvature," Physical Review X {\bf 7}(4), 041042 (2017).
	
	\bibitem{Su} W. P. Su, J. R. Schrieffer, and A. J. Heeger, "Solitons in polyacetylene," Phys. Rev. Lett. {\bf 42}(25), 1698 (1979).
	\bibitem{Asb} J. K. Asb\'oth, L. Oroszlány, and A. Pályi, "The su-schrieffer-heeger (ssh) model," Lect. Notes Phys. {\bf 919}, 1 (2016).
	\bibitem{Li} X. B. Li, W. K. Huang, Y. Y. Lv, K.-W. Zhang, C.-L. Yang, B.-B. Zhang \emph{et al}., "Experimental observation of topological edge states at the surface step edge of the topological insulator ZrTe 5," Phys. Rev. Lett. {\bf116}(17), 176803 (2016).
	\bibitem{Chen} Z. G. Chen, R. Y. Chen, R. D. Zhong, J. Schneeloch, C. Zhang, Y. Huang, F. Qu, R. Yu, Q. Li, G. D. Gu, and N. L. Wang, "Spectroscopic evidence for bulk-band inversion and three-dimensional massive Dirac fermions in ZrTe5," Proc. Nat. Acad. Sci. USA {\bf 114}(5), 816 (2017).
	
	
	
	\bibitem{JB} C. J\"{u}r{\ss} and D. Bauer, "High-harmonic generation in Su-Schrieffer-Heeger chains," Phys. Rev. B {\bf 99}(19), 195428 (2019).
	\bibitem{Ma} C. Ma, X.-B. Bian, and T.-Y. Du, "Role of symmetry breaking in high-order harmonic generation from Su-Schrieffer-Heeger systems," Phys. Rev. B {\bf 106}(12), 125117 (2022).	
	\bibitem{Roth} S. Roth and D. Carroll, \textit{One-Dimensional Metals: Conjugated Polymers, Organic Crystals, Carbon Nanotubes and Graphene}, 3rd ed. (Wiley, New York, 2015).
	\bibitem{Kagoshima} S. Kagoshima, H. Nagasawa, and T. Sambongi, \textit{One Dimensional Conductors}, Springer Series in Solid-State Sciences Vol. 72 (Springer, Berlin, 2012).
	
	
	
	\bibitem{Bao} C. Bao, P. Tang, D. Sun, and S. Zhou, "Light-induced emergent phenomena in 2D materials and topological materials," Nat. Rev. Phys. {\bf 4}(1), 33 (2022).
	\bibitem{Flo_LRB} R.-B. Liu and B.-F. Zhu, "Adiabatic stabilization of excitons in an intense terahertz laser," Phys. Rev. B {\bf 66}(3), 033106 (2002).
	\bibitem{Floquet} G. Floquet, "Sur les équations différentielles linéaires à coefficients périodiques," Sci. Ann. Ec. Norm. Super. 2nd Ser. {\bf 12}, 47 (1883).
	\bibitem{Eckardt} A. Eckardt and E. Anisimovas, "High-frequency approximation for periodically driven quantum systems from a Floquet-space perspective," New J. Phys. {\bf 17}(9), 093039 (2015).
	\bibitem{Rudner3} M. S. Rudner and N. H. Lindner, "Band structure engineering and non-equilibrium dynamics in Floquet topological insulators," Nat. Rev. Phys. {\bf 2}(5), 229 (2020).
	
	
	\bibitem{Yan1} J.-Y. Yan,  "Theory of excitonic high-order sideband generation in semiconductors under a strong terahertz field," Phys. Rev. B {\bf 78}(7), 075204 (2008).
	\bibitem{Yan2} J.-Y. Yan, High-order sideband generation in a semiconductor quantum well driven by two orthogonal terahertz fields," J. Appl. Phys. {\bf 122}(8), 084306 (2017).
	\bibitem{Nagai} K. Nagai, K. Uchida, N. Yoshikawa, T. Endo, Y. Miyata, and K. Tanaka, Commun, "Dynamical symmetry of strongly light-driven electronic system in crystalline solids." Phys. {\bf 3}(1), 137 (2020).
	
	\bibitem{Jin} J.-Z. Jin, H. Liang, X.-R. Xiao, M.-X. Wang, S.-G. Chen, X.-Y. Wu, Q. Gong, and L.-Y. Peng, "Contribution of Floquet-Bloch states to high-order harmonic generation in solids," Phys. Rev. A {\bf 100}(1), 013412 (2019).
	\bibitem{Neufeld} O. Neufeld, D. Podolsky, and O. Cohen,  "Floquet group theory and its application to selection rules in harmonic generation," Nat. Commun. {\bf 10}(1), 405(2019).
	\bibitem{Ikeda} T. N. Ikeda, K. Chinzei, and H. Tsunetsugu, "Floquet-theoretical formulation and analysis of high-order harmonic generation in solids," Phys. Rev. A {\bf 98}(6), 063426 (2018).
	\bibitem{Faisal} F. H. M. Faisal and J. Z. Kami\'nski, "Floquet-Bloch theory of high-harmonic generation in periodic structures," Phys. Rev. A {\bf 56}(1), 748 (1997).
	
	
	
	
	\bibitem{Rana} N. Rana, M. S. Mrudul, D. Kartashov, M. Ivanov, and G. Dixit, "High-harmonic spectroscopy of coherent lattice dynamics in graphene," Phys. Rev. B {\bf 106}(6), 064303 (2022).
	\bibitem{Li2} J. Li, X. Zhang, S. Fu, Y. Feng, B. Hu, and H. Du, "Phase invariance of the semiconductor Bloch equations," Phys. Rev. A {\bf 100}(4), 043404 (2019).
	\bibitem{Vampa2} G. Vampa, C. R. McDonald, G. Orlando, D. D. Klug, P. B. Corkum, and T. Brabec, "Theoretical analysis of high-harmonic generation in solids," Phys. Rev. Lett. {\bf 113}(7), 073901 (2014).
	
	\bibitem{Mrudul}M. S. Mrudul, \'{A}. Jim\'{e}nez-Gal\'{a}n, M. Ivanov, and G. Dixit, "Light-induced valleytronics in pristine graphene," Optica {\bf8}(3), 422 (2021).
	\bibitem{Golde}D. Golde, T. Meier, and S. W. Koch, "High harmonics generated in semiconductor nanostructures by the coupled dynamics of optical inter-and intraband excitations," Phys. Rev. B {\bf77}(7), 075330
	(2008).
	
	\bibitem{zhaoyp} Y.-P. Zhao, G. Wang, S.-J. Ding, and T.-Y. Du,  "Impact of donor and acceptor dopants in high-harmonic generation spectra of solids," J. Opt. Soc. Am. B {\bf 38}(7), 2223 (2021).	
	\bibitem{dtyol} T.-Y. Du, "Observing quantum-path interference and Van Hove singularity in polarization-resolved high-harmonic spectroscopy," Opt. Lett. {\bf 46}(9), 2007 (2021).
	\bibitem{song} L. Song, C. Ma, and T.-Y. Du, "Optimization of surface high-order harmonic generation by tailoring nanostructures," J. Opt. Soc. Am. B {\bf 39}(10), 2678 (2022).
	
	

	

	
	
	
	

	
	
	
	\bibitem{Du1} T.-Y. Du, "Control of high-order harmonic emission in solids via the tailored intraband current," Phys. Rev. A {\bf 104}(6), 063110 (2021).
	\bibitem{WangGan} G. Wang and T.-Y. Du, "Quantum decoherence in high-order harmonic generation from solids," Phys. Rev. A {\bf 103}(6), 063109 (2021).
	\bibitem{Vampa3}G. Vampa, C. R. McDonald, G. Orlando, P. B. Corkum, and
	T. Brabec, "Semiclassical analysis of high harmonic generation in bulk crystals," Phys. Rev. B {\bf 91}(6), 064302 (2015).
	\bibitem{Bielke} L. Bielke, C. J\"{u}r{\ss}, V. Burgtorf, and D. Bauer, "Formation of the solid-state high-order harmonic generation plateau through destructive interference," Phys. Rev. A {\bf 107}(3), 033111 (2023).
	
	
	
	
	
	
\end{thebibliography}
\end{document}